%% file: 00_anchor.tex
\documentclass[journal]{vgtc}                
\ifpdf
  \pdfoutput=1\relax                   
  \usepackage{graphicx}                
  \DeclareGraphicsExtensions{.pdf,.png,.jpg,.jpeg} 
\else
  \usepackage{graphicx}                
  \DeclareGraphicsExtensions{.eps}     
\fi%

\graphicspath{{figures/}{pictures/}{images/}{./}} 

\usepackage{microtype}                 
\PassOptionsToPackage{warn}{textcomp}  
\usepackage{textcomp}                  
\usepackage{mathptmx}                  
\usepackage{times}                     
\usepackage{cite}                      
\usepackage{tabu}                      
\usepackage{booktabs}                  
\usepackage{multirow}
\usepackage{makecell}
\usepackage{enumitem}
\usepackage{graphicx}
\usepackage{caption}

\onlineid{0}

\vgtccategory{Research}
\vgtcpapertype{please specify}

\title{Towards Evaluating Exploratory Model Building Process
    \\with AutoML Systems}

\author{Sungsoo Ray Hong, 
        Sonia Castelo, 
        Vito D'Orazio, 
        Christopher Benthune, 
        \\
        Aecio Santos, 
        Scott Langevin, 
        David Jonker, 
        Enrico Bertini, 
        Juliana Freire
}
\authorfooter{
\item
 Sungsoo Ray Hong is with George Mason University E-mail: shong31@gmu.edu.
 \item
 Sonia Castelo, Aecio Santos, Enrico Bertini, and Juliana Freire are with New York University. E-mail: \{s.castelo, aecio.santos, enrico.bertini, juliana.freire@nyu.edu\}@nyu.edu
 \item
 Vito D'Orazio is with University of Texas at Dallas. E-mail: dorazio@utdallas.edu.
 \item
 Christopher Benthune, Scott Langvin, and David Jonker are with Uncharted Software. E-mail: \{cbethune, slangevin, djonker\}@uncharted.software.
}

\shortauthortitle{Biv \MakeLowercase{\textit{Hong et al.}}: Towards Evaluating Exploratory Model Building Process}

\abstract{The use of Automated Machine Learning (AutoML) systems are highly open-ended and exploratory. While rigorously evaluating how end-users interact with AutoML is crucial, establishing a robust evaluation methodology for such exploratory systems is challenging. First, AutoML is complex, including multiple sub-components that support a variety of sub-tasks for synthesizing ML pipelines, such as data preparation, problem specification, and model generation, making it difficult to yield insights that tell us which components were successful or not. Second, because the usage pattern of AutoML is highly exploratory, it is not possible to rely solely on widely used task efficiency and effectiveness metrics as success metrics. To tackle the challenges in evaluation, we propose an evaluation methodology that (1) guides AutoML builders to divide their AutoML system into multiple sub-system components, and (2) helps them reason about each component through visualization of end-users' behavioral patterns and attitudinal data. We conducted a study to understand when, how, why, and applying our methodology can help builders to better understand their systems and end-users. We recruited 3 teams of professional AutoML builders. The teams prepared their own systems and let 41 end-users use the systems. Using our methodology, we visualized end-users' behavioral and attitudinal data and distributed the results to the teams. We analyzed the results in two directions: what types of novel insights the AutoML builders learned from end-users, and (2) how the evaluation methodology helped the builders to understand workflows and the effectiveness of their systems. Our findings suggest new insights explaining future design opportunities in the AutoML domain as well as how using our methodology helped the builders to determine insights and let them draw concrete directions for improving their systems.
} 


\CCScatlist{ 
 \CCScat{H.5.2}{Information Interfaces And Presentation (e.g., HCI)}%
{User Interfaces}{Evaluation/methodology};
 \CCScat{I.2.5}{Artificial Intelligence}{Program Languages and Software}{Expert system tools and techniques}
}

\vgtcinsertpkg

\begin{document}
\maketitle
\input{sections/01_Introduction}
\input{sections/02_Relatedwork}
\input{sections/03_Research_Methodology}
\input{sections/04_Study1_AutoML}

\input{sections/05_Study2_EvalFrame}
\input{sections/06_Conclusion}


\bibliographystyle{bibs/abbrv-doi}

\bibliography{bibs/references}
\end{document}

%% file: sections/01_Introduction.tex
\section{Introduction}

Automated Machine Learning (AutoML) systems present an interactive environment that helps users to build their model even if they have little knowledge about Machine Learning (ML)~\cite{Yang2018Grounding, hong2019disseminating}.
Commercial AutoML services, such as SageMaker~\cite{AWSSageMaker}, Microsoft ML Studio\cite{MicrosoftAzure}, and Google Cloud AutoML~\cite{GoogleCloudAutoML}, have successfully established large and active user communities. Meanwhile, research communities in Visual Analytics and Human-Computer Interaction have investigated new techniques and systems that can help end-users in building ML models~\cite{talbot2009ensemblematrix, zhang2019manifold, dingen2019regressionexplorer}. Such approaches are largely based on our theoretical understanding of how people explore information and gain insights using interactive systems, including sense-making loop~\cite{pirolli2005sensemaking}, information foraging~\cite{pirolli1999information}, and characterization of tasks and queries in Exploratory Visual Analysis~\cite{thomas2005illuminating, alspaugh2013building, battle2019characterizing}. Understanding how users use AutoML systems to perform their \textit{exploratory model building} in constructing, understanding the resultant models, and recognizing the implications of their model choices is an important topic in visual analytic because the way in which a user builds a ML model is highly exploratory and can be directly supported by visual interfaces~\cite{cashman2019user, hong2019disseminating}. Gaining a deeper understanding of how users think with AutoML systems and where things may fail is crucial for determining how we can better support their ML model building process with visual analytic solutions.


However, our empirical understanding of how end-users perform their exploratory model building using AutoML systems remains limited~\cite{Yang2018Grounding, hong2019disseminating}. For AutoML builders, evaluation of their system is an indispensable part of gaining insights from end-users~\cite{doshi2017towards}. However, such evaluation is not without challenges~\cite{chen2016diagnostic} mainly because of two reasons. One, AutoML systems are \textit{complex}; they include a variety of sub-components built for supporting different sub-tasks, such as viewing data feature distribution, specifying a target metric, or inspecting and comparing ML models. This can hinder the identification of which sub-components are useful and which act as a roadblock~\cite{olson2014ways}. Heuristic evaluation methods, such as NASA TLX~\cite{hart1988development} and System Usability Scale (SUS)~\cite{bangor2008empirical}, have been widely used in evaluating complex systems such as visual analytic systems. However, such a ``quick-and-dirty'' approach \cite{buley2013user} may not be suitable to generate specific insights that builders can leverage to improve their system~\cite{olson2014ways}. Second, the way that end-users interact with AutoML is highly \textit{exploratory} in general; end-users may switch back-and-forth between multiple sub-components in building a ML model~\cite{fails2003interactive, amershi2015modeltracker} without having a clear group-truth success metric in mind, especially for those who don't have ML-related knowledge. Such an aspect makes it difficult to rely solely on conventional evaluation metrics such as task efficiency (e.g., task completion time) or task effectiveness (e.g., model accuracy)~\cite{plaisant2004challenge}. To counterbalance the shortcomings of conventional metrics, novel metrics such as false discovery rate~\cite{zgraggen2018investigating} and interaction rates~\cite{battle2016dynamic} have been proposed in research on Exploratory Visual Analysis (EVA). To date, however, defining the success metrics that can be adopted in real contexts in EVA remains elusive \cite{battle2019characterizing}. 

The complexity in system design and exploratory usage pattern has imposed challenges in evaluating AutoML systems~\cite{hong2019disseminating} and more broadly, EVA tools~\cite{ plaisant2004challenge, battle2019characterizing}. In this work, we propose an evaluation methodology that helps builders to perform a systemic and rigorous evaluation. To tackle the challenges, we designed our methodology based on the following two key directions. \textit{Modular} evaluation: our methodology guides builders to partition their whole system into hierarchically structured sub-components. In this way, builders can semantically break down their unit-of-analysis based on their evaluation needs and more specifically focus on each component's end-user effect. \textit{Multi-faceted} evaluation: our methodology enables builders to collect and leverage multi-faceted aspects of user data comprised of behavioral usage patterns (i.e., how users actually interacted with a system) and attitudinal aspect (i.e., what users were thinking when interacting with a system). We present the data collected through our modular and multi-faceted methodology in a set of visualizations, leading builders to accurately evaluate usage patterns and generate meaningful insights.


To understand how the adoption of our methodology could improve the evaluation of AutoML systems, we collaborated with three teams that develop AutoML systems: Distil~\cite{langevin2018distil}, TwoRavens~\cite{honakerdorazio14, d2018tworavens}, and Visus~\cite{santos2019visus}. The three teams are comprised of 11 professional AutoML builders. Each team applied our methodology to evaluate their system. To evaluate the three systems, we conducted an observational study where we recruited 41 domain experts from the social sciences and tasked them to use one of the three systems to create a regression model and a classification model. Using our evaluation methodology, the three teams collected behavioral and attitudinal data from the participants. We built ``evaluation cards'' that support \textit{within-system analysis}---the analysis that supports sub-component-wise evaluation within a system, and \textit{between-system analysis}---the analysis that helps builders to identify relative strengths and weaknesses of their system with a system-wise comparison. Using our cards, each builder reviewed their system individually and collectively with their group. We collected review outcomes that 11 individuals created and conducted semi-structured interviews with a representative from each team. Then we analyzed (1) what types of insights that AutoML builders were able to discover about their systems, and (2) whether using our evaluation methodology presented unique insights that would be unattainable otherwise. 

This work offers the following contributions.
\begin{itemize}
    \item \textbf{Evaluation Methodology}: We present an evaluation methodology that researchers and practitioners can flexibly customize and adopt to evaluate AutoML systems.
    \item \textbf{Insights about AutoML systems}: We present insights that the three teams identified regarding how their end-users interacted with their systems and what were some common road-blocks. Based on the findings, We briefly discuss design opportunities for AutoML systems.
    \item \textbf{Insights about our evaluation methodology}: We present how the three teams used our evaluation framework, when our methodology presented unique insights, and for which reasons. We also discuss how our methodology can be applied in evaluating different types of exploratory systems in the future.
\end{itemize}

%% file: sections/02_Relatedwork.tex
\section{Related Work}

We review state-of-the-art AutoML approaches in VIS and HCI communities. We then discuss the growing importance of the evaluation in AutoML communities while when existing methodologies can fall short of presenting useful insights when evaluation. We then discuss some key insights for evaluating systems for exploratory analysis.

\subsection{Research in AutoML and Empirical Findings}
The notion of AutoML---an interactive system that support end-users in creating their models regardless of their ML-related knowledge---has been discussed in early seminal work proposed by Ware et al.~\cite{ware2001interactive} and Fails and Olsen~\cite{fails2003interactive}. Ware et al. introduced Visual Decision-tree Constructor~\cite{ware2001interactive}, which leverages a graphical user interface to help statisticians creating decision-tree based models. Fails et al. proposed Interactive Machine-Learning (IML), a concept that presents interaction modalities to humans in stages of building and using ML models so that they can better reflect on their domain knowledge~\cite{fails2003interactive}. With the rise of ML and AI, researchers in HCI and InfoVis have proposed AutoML tools with novel features. For instance, Talbot et al. proposed EnsembleMatrix~\cite{talbot2009ensemblematrix}, a tool that linearly combines multiple models to support non-ML experts. Gestalt is an environment that helps engineers transition back and forth between stages of model implementation and analysis~\cite{patel2010gestalt}. Brooks et al. presented FeatureInsights, an analytic tool that helps users identify useful features for building their models~\cite{brooks2015featureinsight}. With rapid progress in neural network and deep learning, Hohman et al. discuss the role of visual analytic as a tool for building better deep-learning based models \cite{hohman2018visual}. ActiVis~\cite{kahng2017Acti}, GAN Lab~\cite{kahng2018gan}, and Summit~\cite{hohman2019s} fall into such approach.

Meanwhile, another line of research stresses the importance of gaining an empirical understanding from end-users in designing AutoML systems~\cite{hong2020human, hong2019disseminating}. Amerishi et al. discuss the importance of considering both end-user concerns and model designs when creating IML systems~\cite{amershi2011effective}. Several studies found that the design of AutoML and IML systems informed by end-users' usage patterns can facilitate their learning of ML-related concepts~\cite{fails2003interactive, doshi2017towards}. This aspect has been observed from studies conducted for both non-ML experts~\cite{chen2016diagnostic, Yang2018Grounding} as well as ML professionals~\cite{brooks2015featureinsight}. In particular, studies agree that the design of AutoML is closely related to helping a user building their \textit{mental model}---an internal representation that enables them to predict the behaviour of a system and the expected ML models visualized therein\cite{kulesza2015principles, chen2016diagnostic, mohseni2018survey}. Evaluation of AutoML is therefore crucial in determining how to better support users. In general, AutoML evaluations are either controlled studies---to rigorously understand the specific effect of their treatments,---case studies---to elicit general feedback from relevant professionals (but not necessarily ML experts) with in-depth knowledge about the targeted domains, or usability metrics---heuristically developed survey-based scores. AutoML research projects that have applied controlled studies include ModelTracker\cite{amershi2015modeltracker}, FeatureInsights\cite{brooks2015featureinsight}, and Squares\cite{ren2017squares}. In contrast, Confusion Wheel\cite{alsallakh2014visual}, VizML\cite{chen2016diagnostic}, Prospector\cite{krause2016interacting}, RuleMatrix\cite{ming2018rulematrix}, Manifold\cite{zhang2019manifold} have applied case studies or usability metrics.

\subsection{Criticism in Methodologies for Evaluating AutoML}
Designing a reliable evaluation methodology is a notable challenge in AutoML. Researches have questioned whether the methods widely adopted in our practice can validly capture the ``measure of success''~\cite{plaisant2004challenge} which is essential determining a better design. Chen et al. discuss the hardship of selecting the ``right'' research method for evaluating IML systems tools~\cite{chen2016diagnostic}. 

One common approach is determining a single metric in controlled experiments. Researchers put significant effort into the operationalizing the metric that can capture the ``success'' of explorative tasks (e.g., false discovery rate~\cite{zgraggen2018investigating}, interaction rates~\cite{battle2016dynamic}), yet consensus has not been made. Plaisant argue that relying solely on controlled experiments may not allow builders to fully capture the degree to which a particular system can support users in a natural environment~\cite{plaisant2004challenge}. Another concern with controlled studies is that tasks to be examined can be artificial and coverage of possible use cases can be narrow when conditions are rigorously controlled. Doshi-Velez and Kim explain that when humans are involved in the evaluation, the tasks used for the evaluation can become ``simplified''~\cite{doshi-Velez2018consideration} and may not reflect the complexity of ML-related tasks in real scenarios. Ren et al. discuss the difficulties of deterministically arguing one's treatment in a control experiment is always useful in realistic use context, as their findings suggest that useful information and information in investigating ML models is highy context-dependent~\cite{ren2017squares}. Another methodological approach is using a case study or some usability metrics that are heuristically developed, such as NASA TLX~\cite{hart1988development} and System Usability Scale (SUS)~\cite{bangor2008empirical}. However, the general consensus in HCI community is that the insights drawn using such evaluation methodology can be often not specific enough for drawing insights for improving the design~\cite{olson2014ways}. To fill the gap, researchers have dedicated effort into creating visualizations that can better measure the behavior of users engaged in exploratory tasks or empirically understanding how professionals such as data scientists perform ML-related tasks. For example, Battle and Heer discuss the importance of considering ``interaction sequences'' and especially ``behavior graphs''~\cite{battle2019characterizing} in analyzing the quality of using a system in the domain of Exploratory Visual Analysis (EVA). InfoVis workshop related to visualizing user interaction logs identified a series of visualization approaches as well as open challenges~\cite{vuillemot:hal-01535913}. 

While systemic and scientific measurement of users' exploratory analysis behavior is crucial, findings in the existing work suggest that such an attempt entails substantial challenges.

%% file: sections/03_Research_Methodology.tex
\section{Research Methodology}

\begin{figure*}[!t]
\centering
\includegraphics[width=6in]{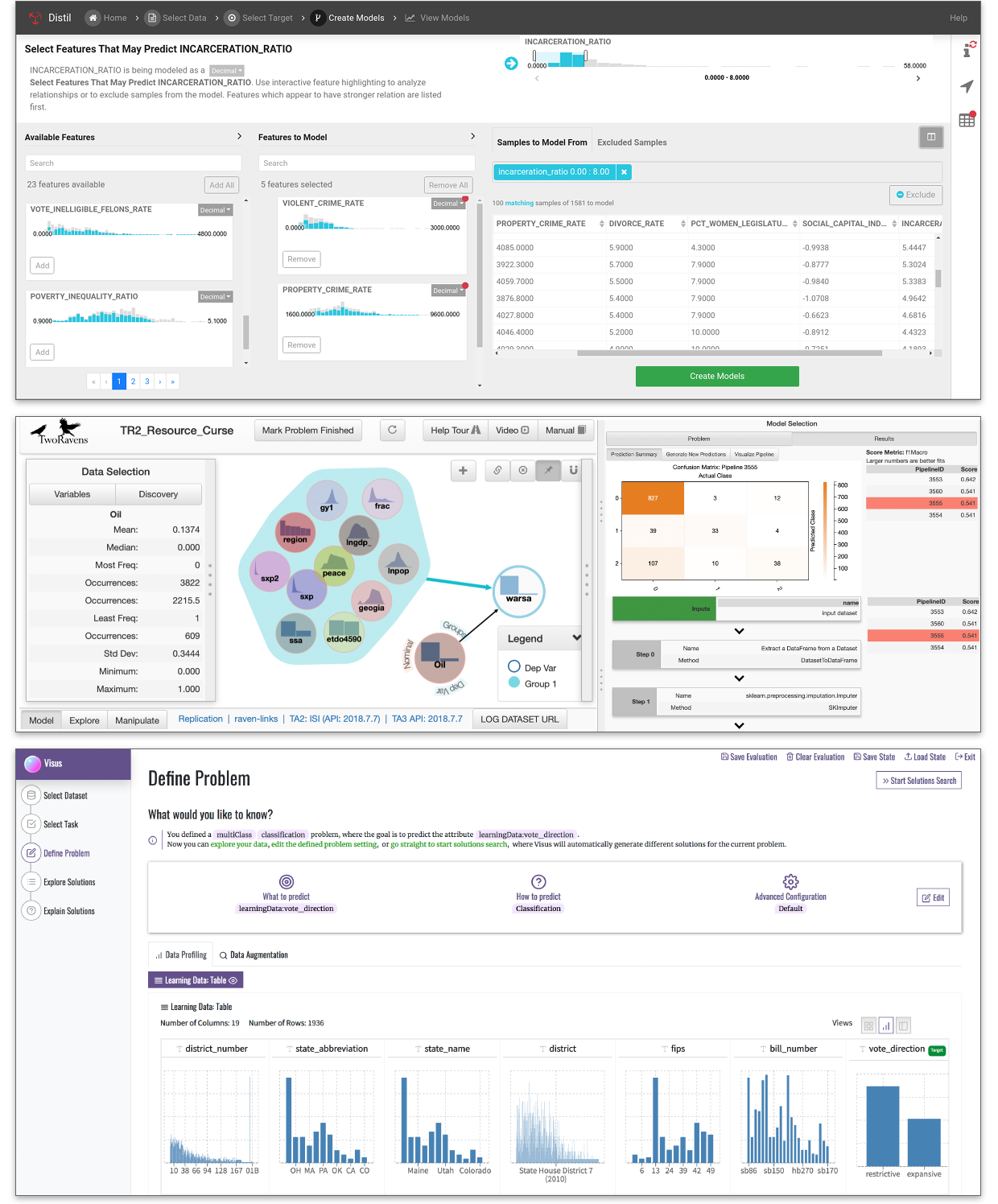}
\caption{Three systems used in evaluation of our evaluation framework: \textbf{Distil} (top), \textbf{TwoRavens} (middle), \textbf{Visus} (bottom)}
\label{fig:systems}
\end{figure*}

Our overarching goal is to develop a methodology that can help AutoML builders can use in evaluation to gain useful insights about how end-users used their system and iron out specific plans for improving their system. In this section, we describe our research process in the following order: how we recruited the three AutoML teams with 11 builders in~\ref{sec:three-automl-systems}, how we developed our evaluation methodology in~\ref{sec:evaluation-methodology}, and how three teams applied our evaluation methodology for evaluation in~\ref{sec:apply-evaluation}.


\subsection{State of the Art AutoML Systems}
\label{sec:three-automl-systems}

We reached out to a consortium comprised of multiple research institutions and working groups. Their mission is to build ``meta machine learning'' systems that would allow researchers (e.g., domain experts) to better discover data-driven insights using state-of-the-art ML and AI technologies. There were several AutoML systems under development in this consortium, and one of their interests was having a formal methodology that they can use for measuring the degree to which their system supports an end-user's exploratory ML model building process. Several teams in a consortium showed their interest in our research goal, and three of the teams, comparised of 11 AutoML builders, decided to collaborate with us: they were responsible for building systems named \textit{Distil}\cite{langevin2018distil} (Fig.~\ref{fig:systems}, top), \textit{TwoRavens}\cite{honakerdorazio14, d2018tworavens}  (Fig.~\ref{fig:systems}, middle), and \textit{Visus}\cite{santos2019visus} (Fig.~\ref{fig:systems}, bottom) respectively. While the three teams had their unique sub-components in their system built for supporting their end-users in a particular way, they shared the common general goal---helping end-users to build a better ML model with less effort. The three teams agreed on improving their own system based on their own design directions for three months before conducting an evaluation using our methodology.

\subsection{Evaluation Methodology}
\label{sec:evaluation-methodology}

Based on the literature review, we determined the following two design directions in shaping our evaluation methodology. 

\vspace{2mm}

\textbf{Modular Component Definition}: The first notable aspect that we identified through the review is that there seem to be trade-offs between the \textit{coverage of evaluation} (e.g., how ``many'' of sub-system components and design factors one aims at evaluating in a single round) and the \textit{quality of findings} (e.g., the depth or rigor) one can expect from using a certain type of evaluation methodology. To strike the balances between the coverage of evaluation and the quality of findings, we define a semi-structured method that allows AutoML builders to \textit{partition} their whole system into a set of sub-components that will be used as a unit-of-analysis in evaluation. We expect that scoping the unit of analysis through a set of components will help builders flexibly define the granularity of their findings, ensuring that any insights gained are not trivial or overly generic.

\textbf{Multifaceted Capturing}: Another insight we found in the review is that AutoML builders might still encounter hardship in gaining meaningful insights with a single ``success metric''; rather, considering \textit{multifaceted} aspects of users' usage of their system might present more comprehensive insights. Therefore, we present a method that AutoML builders can follow to capture behavioral and attitudinal aspects of users' system usage, which are the two pillar of measures widely used in HCI research~\cite{albert2013measuring}. By combining the two perspectives, we expect that builders will be able to gain deeper insights that help determine design insights more reliably than general evaluation methods.

\subsubsection{Modular Component Definition}
Out evaluation methodology allows builders to splite their whole system into a set of sub-components. To do so, we created a reference table that lists possible functionalities that end-users can commonly carry out in AutoML systems. 

Cashman et al.'s workflow concept for Exploratory Model Analysis \cite{cashman2019user} presents the types of activities that a \textit{system} would provide to a user (e.g., ``the \textit{system} uses interactive visualization to provide an initial data overview''). In defining the table, we converted their tasks  based on what a \textit{user} can ask (e.g., ``A \textit{user} looks up a dataset''). This conversion was necessary because some steps in the workflow are not directly related to a user's action (e.g., step 4 ``an automated ML system trains and generates candidate models''). We also added other functionalities increasingly introduced by state-of-the-art AutoML systems, such as data augmentation\cite{gil2019towards, santos2019visus}. After we finalize listing every functionality, we re-structured our functionalities using two-level hierarchy. Structuring the functionalities to two-level hierarchy would allow builders to semantically divide the high-level functions in the first-level then more specifically define a target component-of-evaluation in the second level. We reviewed the level-one and level-two functionalities with 11 builders across the three team. Specific functionalities we defined through this process are shown in Table~\ref{table:two-level-specs}. The level-one represents high-level AutoML user goals: (1) preparing \textit{data}, (2) defining a \textit{problem} based on the chosen dataset, and then (3) building a \textit{model} that can solve the specified problem. The level-two represents more specific sub-system features associated with the level-one goals. For example, there are four level-two functionalities associated with the ``Model'': ``summarizing models'', ``explaining a model'', ``comparing models'', and ``export a model''. 

\begin{table}[t]
\caption{Two-level functionalities used in our methodology.}
\label{table:two-level-specs}
\footnotesize
\bgroup
\def\arraystretch{1.3} 
\begin{tabular}{lll}
\hline
\multicolumn{2}{c}{\textbf{Functionalities}}                                       & \multicolumn{1}{c}{\multirow{2}{*}{\textbf{User activity description}}}                                                                                     \\ \cline{1-2}
\multicolumn{1}{c}{\textbf{Level 1}}     & \multicolumn{1}{c}{\textbf{Level 2}} & \multicolumn{1}{c}{}                                                                                                                                        \\ \cline{1-3}
\multirow{4}{*}{\textbf{Data}} & Open a dataset                        & A user selects a dataset                                                                                                                                    \\
                                           & Explore a dataset                     & \makecell[cl]{A user looks up a dataset (e.g.,\\ check the distribution of a\\ feature, see a specific instance)}                                                              \\
                                           & Augment a dataset                     & \makecell[cl]{ A user augments a dataset\\ (e.g., a user searches other\\ relevant datasets and joins new\\ features with the chosen dataset)\\ or adds/removes features}            \\
                                           & Transform a dataset                   & \makecell[cl]{A user cleans/bins features\\ in a dataset}                                                                                                               \\ \hline
\textbf{Problem}                & Specify a problem                     & \makecell[cl]{A user specifies a series of\\ parameters required for an\\ AutoML system to generate\\ models (i.e., target metric, type\\ of ML models, advanced\\ settings, etc.)} \\ \hline
\multirow{4}{*}{\textbf{Model}}  & Summarize models                      & \makecell[cl]{A user requests/looks up\\ general information about a\\ set of models generated by an\\ AutoML system}                                                            \\
                                           & Explain a model                       & \makecell[cl]{A user views detailed\\ information about a model\\ (e.g., performance, cases for\\ making accurate or inaccurate\\ predictions)}                                        \\
                                           & Compare models                        & \makecell[cl]{A user requests information to\\ compare multiple models}                                                                                                      \\
                                           & Export a model                        & A user exports a model                                                                                                                                     \\  \hline
\end{tabular}
\egroup
\end{table}

\begin{table}[b!]
\caption{Component definition for Visus extends the two-level functionalities table to three levels by dividing the system's sub-components.}
\label{table:two-level-specs-for-visus}
\footnotesize
\bgroup
\def\arraystretch{1.3} 
\begin{tabular}{lll}
\hline
\textbf{Level 1}     & \textbf{Level 2} & \textbf{Visus Component}\\ \hline
\multirow{4}{*}{\textbf{Data}}          & Open a dataset         & Open a dataset \\
                                        & Explore a dataset      & Explore a dataset                                                              \\
                                        & Augment a dataset      & N/A \\
                                        & Transform a dataset    & N/A \\ \hline
\multirow{3}{*}{\textbf{Problem}}    & \multirow{3}{*}{Specify a problem}   & Select a target metric                                                                                                                           \\
                                     &                                      & Define a problem type                                                                                                                            \\
                                     &                                      & Advanced configurations                                                                                                                          \\ \hline
\multirow{7}{*}{\textbf{Model}}      & Summarize models                     & N/A                                                                                                                                              \\
                                     & \multirow{4}{*}{Explain a model}     & \makecell[cl]{See a confusion matrix\\ (for classification models)}                                                                                               \\
                                     &                                      & \makecell[cl]{See confusion scatter plot\\ (for regression models)}                                                                                               \\
                                     &                                      & \makecell[cl]{See partial dependency plots\\ (for regression models)}                                                                                             \\
                                     &                                      & \makecell[cl]{See partial dependency plots\\ (for regression models)}                                                                                             \\
                                     & Compare models                       & Compare models                                                                                                                                   \\
                                     & Export a model                       & Export a model      \\  \hline
\end{tabular}
\egroup
\end{table}

Our evaluation framework suggests builders going over each level-two functionality and choose one of the following actions: (1) \textit{Applying}: If the scope of a second-level functionality is suitable as a unit-of-analysis in evaluation, builders may adopt it without modification. In this case, builders will have a single component corresponding to the second-level component, (2) \textit{Subdividing}: If the scope of a level-two functionality is overly broad, builders can subdivide the level-two to multiple level-three components for more fine measurement. (3) \textit{Dropping-out}: If a second-level functionality is not relevant to a system, builders may omit it. (4) \textit{Creating}: Finally, if a builder cannot find relevant level-two functionalities, the builder can create a new component in level-two. Table~\ref{table:two-level-specs-for-visus} shows how builders can apply the reference table in defining their own sub-components. This example is the table made by Visus team for this study. They applied ``Explore a Dataset'' without modification; omitted ``Augment a dataset'' and ``Transform a dataset'' as there were no relevant system features associated with these functionalities; and subdivided ``Explain a model'' into four components: ``See a confusion matrix'', ``See rule matrix'', ``See confusion scatter plot'', and ``See partial dependency plots''. 

\subsubsection{Multifaceted End-user Behavior \& Perception Capture}
The second step of applying our evaluation method is to define what aspects of data related to end-users that a builder would measure, which we elaborate as follows:

\textbf{Behavioral data}: Behavioral data captures \textit{what end-users actually did}. For example, we can explain (1) how much time an end-user spent on each sub-component and (2) how many times (s)he visited sub-components. With behavioral logs, builders can also identify a user's exploration pattern---whether their workflow was \textit{linear} (i.e., a user built a model without using components back-and-forth) or \textit{back-and-forth} (i.e., a user moved back and forth in using multiple components to build their model). Table~\ref{table:log} lists attributes our that log suggests. We recommend that additional data be collected for components linked to certain functionalities (see ``Other'' in Table~\ref{table:log}. For instance, for any sub-components related to ``Specify a Problem'', the log suggests saving a set of parameters an end-user sent to a system, such as a target metric, parameters for advanced configurations. For any components associated with ``Explain a model'', the log suggests collecting specific information about the model that the user viewed. In ``Export a model'', the final performance of a model that an end-user built can be logged. However, in our the study (described in Section 3.4), we chose not to record model performance, as model performance can vary depending on how end-users specify their problem---for example, selecting different performance metrics such as F1 score or model accuracy. 

\textbf{Attitudinal Data}: Attitudinal data reveals \textit{what end-users felt} while engaging with a system. In collecting this data, we first collect an end-user's \textit{per-sub-component perception} about perceived efficiency and effectiveness 
based on prior work (e.g., Albert and Tullis's approach~\cite{albert2013measuring}). In asking end-users' perception, each team prepared a screenshot of every level-two component. In eliciting end-users' perception, we asked users to respond with the 5-level Likert scale, where the lowest score is 1 (``Very Hard'') and the highest is 5 (``Very Easy'') (following practice used in NASA-TLX~\cite{hart1988development}). Some example questions given to the end-users were as follows (the questions below assume the screenshot of the component is located on the left and the target task for the component is ``set a target metric to predict''):
\begin{enumerate}
    \item \textbf{Perceived efficiency}: ``Using the component on the left, how hard did you have to mentally and/or physically work to \textit{set a target metric to predict}?''
    \item \textbf{Perceived effectiveness}: ``Using the component on the left, how hard or easy was it for you to accomplish \textit{setting a target metric to predict}?''
\end{enumerate}
To captures end users' general perception about a system, we additionally applied 10 questions proposed in SUS~\cite{brooke1996SUS}, which together yield a usability score between 0 and 100.

\begin{table}[t]
\caption{User log specification}
\label{table:log}
\footnotesize
\bgroup
\def\arraystretch{1.3} 
    \begin{tabular}{p{2.1cm} p{5.5cm}}
            \hline
            \textbf{Attribute name} & \textbf{Description} \\ \hline
            \textbf{Time stamp} & An ISO-8601-formatted timestamp that shows when a user used one of the UIs included in a particular component \\ \hline
            \textbf{LV1 id} & Lv.1 identifier associated with a component (e.g., Data) \\ \hline
            \textbf{LV2 id} & Lv.2 identifier associated with a component (e.g., Explain a model) \\ \hline
            \textbf{Comp id} & Component identifier (e.g., See PDPs) \\ \hline
            \textbf{Other} & Only available when using components related to ``Specify a problem'' (a parameters a user set for creating models) or ``Explain a model'' (a set of information about a model a user looked through) \\ \hline
    \end{tabular}
\egroup
\end{table}

\begin{figure*}[!t]
\centering
\includegraphics[width=7.1in]{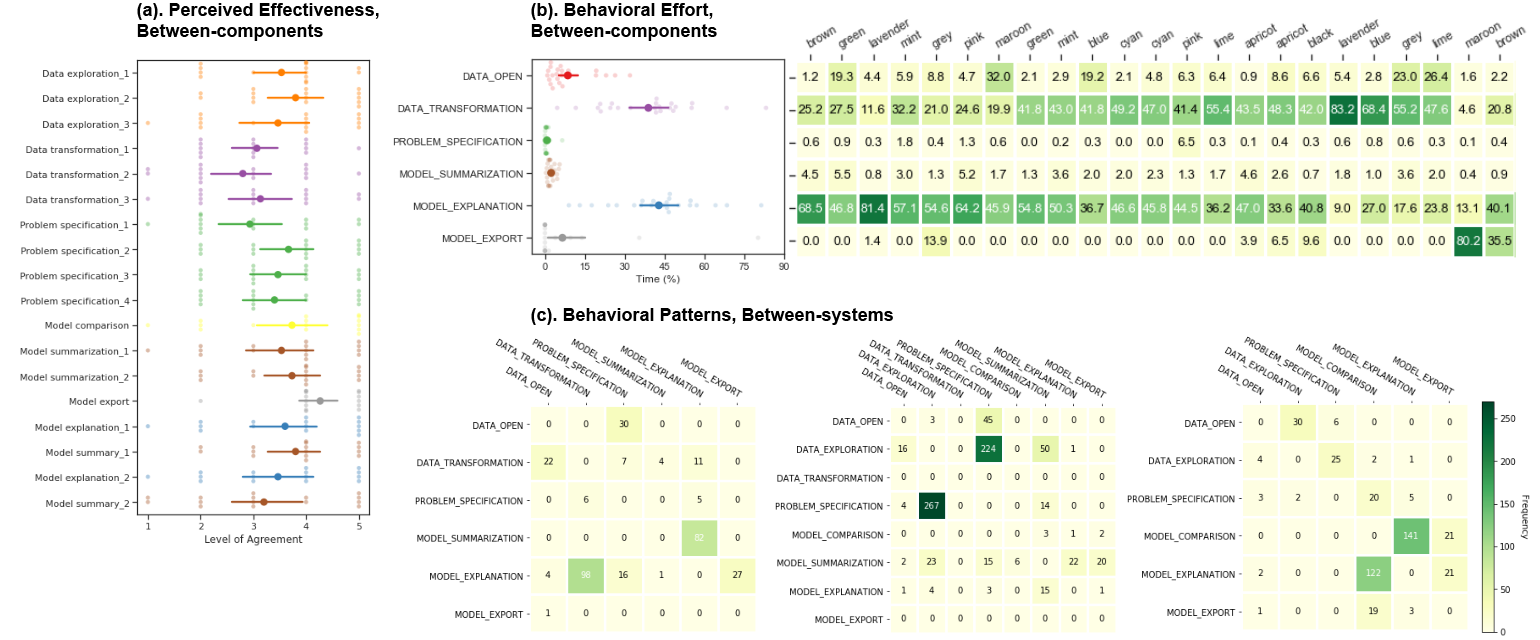}
\captionsetup{justification=centering}
\caption{Evaluation cards: (a) a visualization of perceived effectiveness between components, TwoRavens. (b) A visualization of behavioral effort between components, Distil, (c) A visualization of behavioral patterns between three systems, Distil (left), TwoRavens (middle), and Visus (right)}
\label{fig:cards}
\end{figure*}

\subsubsection{Creating Evaluation Cards}
\label{sec:evaluation-cards}
The last step of our methodology is creating a material for evaluation. Specifically, our methodology creates ``Evaluation Cards'' that contain a set of static visualizations that shows behavioral and attitudinal data. Such decision was made to minimize a learning curve. 

The design of the evaluation cards is the product of a collaborative effort by a team of 3 visualization researchers and 7 AutoML software engineers (3 of them are the team representatives of Distil, TwoRavens, and Visus). An initial design has been built by the visualization researchers. Following a process of user-centered design (UCD), the researchers presented their initial outcomes to AutoML builders. AutoML engineers reviewed the outcomes, discussed the design in group, and reported issues. Meanwhile, three representatives shared the design to their team to check technical feasibility and collect the team's feedback. Based on the received feedback, visualization researchers revised the visualization and prepared the improved version for next review. The team conducted four rounds of review and revision throughout a period of 8 weeks. Consequently, we came up with the cards with the four visualization categories as follows:

\begin{enumerate}
    \item \textbf{Descriptive Results}: This category visualizes the overall behavioral performance of end-users and their perceived score about the whole system. The metrics used in this category are: task-completion time, interaction steps required for task-completion, and SUS score. 
    \item \textbf{Attitudinal}: This category shows end-users' perceived efficiency and effectiveness of every sub-component. Box-and-Whisker charts are used to visualze 5-level likert scale results collected from a group of end-users. Fig.~\ref{fig:cards} (a) shows an example of plots for showing perceived efficiency of TwoRavens. 
    \item \textbf{Behavioral 1--User Effort}: This category presents how much of time end-users used in each of sub-components. Fig.~\ref{fig:cards} (b) shows an example of Distil. Box plots on strip chats in the left hand side shows the relative time spent per sub-component (e.g., Distil end-users tend to spend a longer time in "data transformation" and "model explanation" than other sub-components). The right-hand side shows the raw information; each column shows each user's allocation of time. 
    \item \textbf{Behavoiral 2--Exploration Pattern}: This category shows exploration pattern of end-users using from-to matrix (see Fig.~\ref{fig:cards} (c)). The rows in the matrix indicate the ``from'' sub-components while the columns show ``to'' components, and color indicate the frequency of switching between components. For instance, while the switching patterns in Visus (right) are ``linear'' (i.e., end-users built their model step-by step without moving back-and-forth except model explanation and model comparison), we can see that the patterns in Distil (left) seems more exploratory.
\end{enumerate}

Each category has two sections: within-system analysis, and between-system analysis. In total, Our evaluation methodology presents Evaluation Cards comprised of 8 sections for one AutoML system.
\begin{enumerate}
    \item \textbf{Within-system Analysis}: This section helps builders examining their own system by supporting between-component analysis; how end-users' usage patterns and perceived quality vary across different sub-components? For example, Fig.~\ref{fig:cards} (a) and (b) present sub-component-wise visualization for TwoRavens and Distil respectively. Some typical insights we intend to present through this section are: (1) Which sub-components blocked end-users' model building task? (2) Which sub-components end-users felt useful or not useful? (3) How much of time end-users spent in each of sub-components? (4) What are the common sub-component exploration patterns (i.e., step-by-step or highly back-and-forth)?
    \item \textbf{Between-system Analysis}: This section helps builders benchmark their systems with other systems and examine their relative strengths and weaknesses. While each system has its specific features implemented for supporting slightly different tasks, those systems share some common features. We intend to help AutoML builders identifying which sub-components of their system are relatively weaker than those in other systems, and what they can learn from other systems to improve their weaknesses.
\end{enumerate}

\subsection{Applying Evaluation Framework to Three Systems}
\label{sec:apply-evaluation}
We worked with the three AutoML teams to apply our evaluation methodology in their systems. Using Table~\ref{table:two-level-specs}, Distil defined 6 terminal sub-components, TwoRavens had 18, and Visus yielded 11 (see ``Visus Component'' in Table~\ref{table:two-level-specs-for-visus}). Then three teams implemented end-user interaction track log using Table~\ref{table:log}. Three teams made them ready for collecting data from end-users and creating evaluation cards. We and the 11 AutoML builders across the three teams had the following objectives, which we will deliver in section 4 and 5 respectively:
\begin{enumerate}
    \item \textbf{Understanding AutoML systems}: Our first goal is to evaluate our three systems and gain a deeper understanding of common road block in AutoML and future research directions that could lead to improving the way AutoML support end-users in building reliable, trustworthy, and high-performing ML models.
    \item \textbf{Evaluating Evaluation Methodology}: Our second goal is to understand whether our evaluation framework can improve the way we evaluate AutoML systems, and if so, why and when.
\end{enumerate}

%% file: sections/04_Study1_AutoML.tex
\section{Study 1. Evaluation of Three AutoML Systems}

We describe the lessons the three teams learned through using our evaluation methodology. We start by describing how we collected data and how we analyzed the data (\ref{sec:systems-eval-methodology}). Then, we report main insights learned by the 11 AutoML builders across the three teams (\ref{sec:systems-eval-results}). Finally, we discuss insights that can extend to more broader AutoML systems based on our meta-analysis of the collected data (\ref{sec:systems-eval-discussion}).

\subsection{Methodology}
\label{sec:systems-eval-methodology}
\subsubsection{Collecting data from social science domain experts}
In evaluating three systems, we prepared two common tasks in AutoML: building a classification model another a regression model. To prepare two tasks, we prepared two datasets from peer-reviewed social science publications; Avery and Fine's dataset on legislator votes for a state immigration policy~\cite{avery2019unpacking} were used for a classification problem. Hawes's dataset~\cite{hawes2017social} was used for a regression model building task where the target variable was the incarceration rate in American states. 

To conduct our study, we recruited 41 participants from a public university in the US (IRB 20MR0003). Because domain experts were the three system's primary user group, we aimed at recruiting graduate and upper-level undergraduate social science majors with experience in analyzing social science datasets but little or no knowledge in ML. Overall, we were successful in recruiting 41 participants that fit our criteria. None of the participants had experience with any of the three AutoML systems. Of the 41 participants, 33 reported they are familiar with social science data analysis. In total, we had 17 females, 23 males, and 1 other. Among our participants, 21 have taken graduate level social science classes, 12 of whom had received a graduate level degree. The ages ranged from 18 to 65, with the median age being 25. 

Participants were randomly assigned to one of three rooms in our lab, with each room corresponding to one of the three AutoML systems. TwoRavens, Distil, and Visus had 15, 13, and 13 participants respectively. Participants were trained for 45 minutes on how to use the assigned system. In the training phase, a representative for each AutoML team first explained the general workflow of the system. Then we let participants freely use the system and ask questions. Training was conducted on a small dataset that contained information on professional baseball players, with the target feature being whether the player made it into the Hall of Fame (classification) or how many of home runs a player would hit (regression). 

Following the training session, each participant was asked to analyze the two social science datasets and solve a defined problem for each. To counterbalance the ordering effect, half of the participants started with the classification dataset and the rest started with the regression dataset. In this phase, the participants were provided a handout that contained a description of the data and the task. For each task, they were instructed to explore the data, build models to predict a designated target variable, and ``export'' the model in which they had the most confidence. In total, participants were given two hours---including the 45-minute training session---to complete the tasks and take our survey. We tried to limit incentives to finish early by providing dinner and soft drinks at the two-hour mark. After completing the two data tasks, each participant completed a survey hosted on Qualtrics. The survey was split into three parts: questions related SUS questions~\cite{brooke1996SUS}, per-sub-component questions, and background information questions. For all the questions we deployed in the survey, we used a five-point Likert scale. In total, the study with our participants resulted in 82 behavioral log files (from 2 task types completed by 41 users) and attitudinal data composed of 41 SUS scores and 41 per-component survey results. 

\subsubsection{Collecting data from AutoML builders}

Using the data we collected from 41 domain experts, we created three sets of evaluation cards for each of the teams. Each team used our evaluation cards to review their system. In performing the review, each team assigned a team organizer who is responsible for communicating with team and taking care of outcomes throughout the review. 

First, team organizers distributed evaluation cards to their team members and asked team members to perform an (1) \textit{individual review}, which aimed at helping help each team member to identify some \textit{insights} they think interesting or useful using evaluation cards. 
In the second step, group organizers performed a (2) \textit{group review} workshop with every team member. The purpose of the workshop is to derive a set of specific \textit{actions} they would work on to improve their system. After the workshop, the group organizer wrote their group review report. Then we conducted a one-hour of (3) \textit{semi-structured interview} with the three organizers to understand how they used our evaluation methodology, when and why they felt evaluation was helpful, and what aspects can be improved to better support an evaluation.
Every interview was transcribed by professional transcribers. In total, we collected Eleven \textbf{individual reviews} that lists up some insights each AutoML builder learned through our evaluation cards, three \textbf{group reviews} that shows action items that each team derived for improving their system, and three \textbf{interview transcriptions}.

\subsubsection{Qualitative analysis process}
Among the three data sources we collected, we used 11 individual reviews and 3 group reviews to discover what insights AutoML builders found using our evaluation methodology. In performing our analysis, we strictly followed an \textit{iterative qualitative coding} process~\cite{saldana2015coding}. We first alternate coding (tagging specific text segments with codes) and analytic memo writing (collecting ideas and insights using tags for generating themes) separately. Two authors went through every review and interview and assigned an code. Next, they shared their code to identify commonalities and discrepancies, discussed continuously until they reached an agreement. Finally, the two authors reviewed all our codes, memos, and themes to build the structure with relevant details. They finalized the structure by diagramming (building the structure of themes that span notable codes). 

\subsection{Results}
\label{sec:systems-eval-results}

\begin{figure}[t]
\includegraphics[width=3.3in]{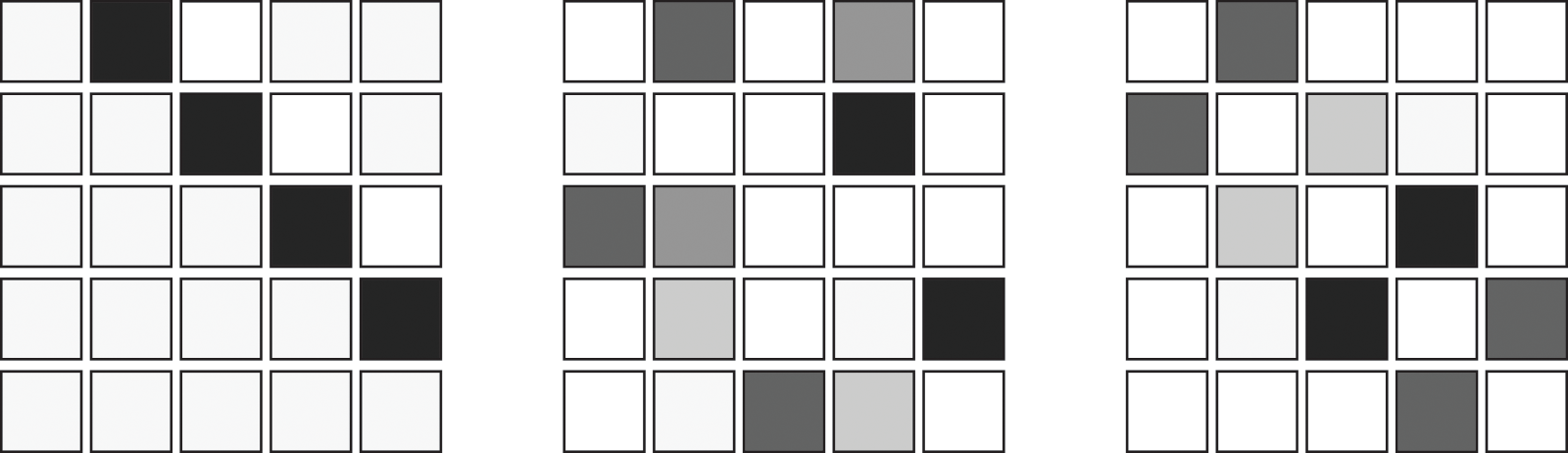}
\centering
\caption{Notable exploration patterns in from-to matrix (rows: from sub-components, columns: to sub-components), linear exploration (left), non-linear exploration (middle), between-component iterations (right)}
\label{fig:fromto}
\end{figure}

%
Our qualitative analysis revealed several recurrent themes about AutoML builders' discovery. One of the most prominent themes is about end-users' system usage patterns that the builders identified. Many participants mentioned that end-users' \textit{linearity} in exploring AutoML systems most interesting. When AutoML builders applying our evaluation methodology to their systems, they assumed that there is a certain stage that an end-user must go through in order; preparing data, specifying a problem, and build a model. The builders determined ``linear'' end-users tend to follow such steps without doing a lot of going back-and-forth across sub-components whereas other ``non-linear'' people jumped around multiple sub-components without an order in their mind. The majority of participants (P1, P2, P3, P4, P5, P6, P7, P8) mentioned that they could use a from-to matrix in Behavioral 2 to see if the patterns drawn in the matrix match with something they expected. We were able to observe three teams  had a deep discussion regarding how different design of the system can influence an end-users' exploration linearity in their group reviews. P7 noted that  \textit{``systems are data-rich and bring together complex elements and components to form coherent workflows. They vary on the axis of directive (linear) vs. non-directive (free-flowing) with Visus and TwoRavens being on opposite ends.''} He went further and affirmed that  \textit{``Each system may benefit by moving closer to the center of that [linearity] axis.''}. These usage patterns regarding linearity can be clearly shown when comparing different heat maps in Figure~\ref{fig:fromto} left and middle. This discussion lead P6 speculating about the possibility of deriving the system-usage linearity as a quantified term and determine how different degree of exploration linearity can effect other end-user behavioral and attitudinal measures.

The evaluation results also revealed novel and unexpected usage patterns about \textit{iterations} between-components (see Figure~\ref{fig:fromto}, right). P1, a builder in Distil, described his unexpected discovery between data exploration and problem specification: \textit{"The usage pattern, which is heavy toggling between data exploration and [problem specification] was a novel discovery and very useful."} Such patterns led P7, a builder in TwoRavens, to think about how to improve their system: \textit{``We should create an easier workflow and transition between data exploration and problem specification.''}. While the builders found such iterations were well-received by end-users, the usage patterns observed in this evaluation revealed a lack of model refinement iterations for some part of their systems. Participants from every team (P2, P3, P9, P11) noticed such a lack of such transitions. This led them to think about possible ways to sidestep this issue. P2 described that the \textit{``Usage pattern shows users moving fairly linearly through the application, with some iteration between model explanation and data transformation, although not as much of that iteration occurred as we hoped - could point to the application not providing enough hints to encourage refinement of models.''} Similarly, P3 noted that one of the possible reasons for observing ``\textit{[...] few transitions to previous states in the workflow [...]}'' was that maybe the lack of visual affordance; ``\textit{it was not clear to users that they could go back to previous states [...]}.'' We found builders considered a lack of iteration in model refinement sub-component a highly important problem, as such usage pattern may imply that their system cannot fully facilitate  end-users' exploratory model building process. PN noted: \textit{``[...] the lower than expected iteration over the data transformation and model explanation screen points to us needing to do a better job of encouraging the user to take additional steps to improve their models.''}

As AutoML systems evolve and attempt to lift the burden of end-users, the importance of visual interfaces for model selection, which includes target tasks such as model summarization, model comparison, has been increasingly growing while adding more complexity. Our evaluation results shed light on the user's perception of model selection sub-components, which generally received low scores across all three systems. The end-users' perceived scores coupled with model selection in the Attitudinal category allowed system builders to drill down how the series of sub-components in model selection drops the overall end-user satisfaction. For instance, P9 felt that sub-components associated with model explanation were perceived as lacking in clarity. P6 noticed that \textit{``Even the confusion matrix, one of the most common explanation methods, is not well received by some people. Rule Matrix is below 3 in terms of their perceived efficiency and effectiveness. Those two need some more easy-to-understand design.''}, suggesting that even classic visualizations may be hard to understand. In a similar vein, P3 identified that the \textit{``low scores for model explanation sub-components seem to be disproportionately negatively impacted by a single component.''} The component was an interactive techniques devised for helping users to inspect ML models based on rules. The Visus builders identified that the explanation method was rated with the worst score, which suggests that it can be complex and hard to interpret by people with less or no ML background.

\subsection{Implications for Design}
\label{sec:systems-eval-discussion}

Based on our meta-analysis of collected data, we derived the four prominent insights for the better design of AutoML systems.

One, builders should design their systems to encourage end-users to operate in iterative, non-linear workflows. AutoML systems are open-ended and exploratory, and the expectation is that end-users would iterate between sub-components as they learn more about the intricacies of the data and the problem they are trying to solve rather than go directly from data to model. We found it is challenging for AutoML systems to simultaneously encourage exploration and to reduce the amount of effort that users would put to build a quality model. Each team discussed this trade-off in their group reviews. It can be challenging to encourage end-users to explore data with an open mind, a fundamental feature of exploratory data analysis \cite{gelman2004exploratory, tukey1977exploratory}. It is not uncommon to have a mental model, to use the data to represent that model as best as possible, and then to spend one's time assessing how well that model fits the data. Even in this case, which is perhaps closer to confirmatory data analysis than exploratory, end-users should be encouraged to use the system to assess the robustness of the mental model in a more holistic fashion.

Two, visualizations should be implemented to allow for interpretation by domain experts who may have little knowledge about visualizations used in data science. Since the target end-users can have little knowledge about ML, builders should be mindful that many visualizations can fall short on facilitating their reasoning about the model's behavior. For example, some end-users reported their trouble in using the confusion matrix, something that is basic in data science community. Along these lines we suggest two. One, builders should have clearly defined plot axis, labels, titles, and captions where necessary. While this is basic, it is crucial that end-users understand the core attributes of the visualizations, and unfortunately it is easy for these elements to be overlooked. Two, builders should provide an example interpretation. We found giving them some example with one or two sentences effective. 
This is not to suggest that builders shy away from complex visualizations. Many complex visualizations are still informative, and domain experts would seek to understand the graphic even if complex. That said, the same applies: clearly defined plot elements for descriptive purposes, and a sample interpretation to aid in drawing inferences.

Three, data manipulations are important for model building, but present usability challenges. We found domain experts are highly interested in transforming the data and engineering features-of-interest even before they specify a problem. However, such functionalities are often not availabe or difficult to use. Data transformations and feature engineering are challenging to implement, as it involves changes that propagate through the system. For example, consider a case where the end-user has built a model, and now wishes to subset the data and build another model. Subsetting the data is a simple transformation, but data summaries require updates and more importantly the user must know the status of the data. Yet, transformations and feature engineering are core elements of model building which we identify better support for the future AutoML design.

Finally, end-user usage patterns differ among different systems yet such patterns may heavily influence on the overall effectiveness of the system. Builders should consider how their system's usage patterns inhibit successful exploratory model building. All three systems had different usage patterns, with some components being more heavily used than others. While there are many reasons for the different rates of use, ranging from the overall system design to the appeal and effectiveness of the individual component, it is important for builders to consider the ways in which usage patterns impact the overall system objective. For example, the appeal of a component may be much greater than other components, leading end-users to spend more time there. Yet, this single component may have limited utility. The opposite is true as well: poorly designed components may be extremely useful but see less use.  Thus, it is important for builders to understand the usage patterns in order to understand the overall effectiveness of system in encouraging end-users to produce high-quality, predictive models.

%% file: sections/05_Study2_EvalFrame.tex
\section{Study 2. Evaluation of Evaluation Methodology}
The focus on the Study 2 is to understand how AutoML builders used our evaluation methodology, where they identified some specific aspects of our methodology were useful/not useful and why, and what were the shortcomings. To understand these questions, the two authors worked to conduct qualitative analysis using three data sources--11 individual reviews, 3 group reviews, and 3 interview transcriptions. The two authors followed the same methodology as Study 1.

\subsection{Results}

\textbf{Effect of modular measurement}: Three team organizers (P2: Distil, P4: Visus, P10: TwoRavens) all mentioned that the main advantages of our evaluation methodology were the support for a modular and hierarchical measure of their system. P10 stated that such modular design presents some framework that forces their team to think about the effect of design in a hierarchical manner which leads them to novel insights that cannot be gained otherwise: \textit{``Thinking about the system at the L1 level, L2 level, and L3 level...was something that I found throughout the entire process to lead to insights, because when you're developing the system, you think a little bit about different levels of analysis, but this really forced you to think about the different levels of analysis''}.  So while the hierarchy was designed to facilitate evaluation leading to insight, the task of mapping to the hierarchy itself also seemed to be instructive in this case. Also, P2 indicated that the defined modular structure allowed for commonalities to be found across the 3 systems, which facilitated meaningful comparisons between systems which they found difficult in evaluating complex systems: \textit{``if you don't break things down and try and have some kind of categorization, you can't build this mental model of the types of things that are done in the applications that are similar...a lot of the value isn't going to be there''}. P4 agreed, indicating \textit{``the main benefit of (modular evaluation) is you can simplify the comparison''}. In general, three teams all mentioned that they appreciated having the capability of flexibly scoping unit-of-analysis by themselves in evaluation.

Our methodology provided behavioral measures for time spent between sub-components (Behavioral 1) and the number of interactions and their switching patterns between different sub-components (Behavioral 2). We found that these categories helped builders to determine insights combining with their modular assessments. Using the top-level measurements in Behavioral 1 and 2, the builders were able to determine broad trends in application behavior. For example, participants observed that users completed their experiment tasks more quickly than expected, and with fewer interactions than thought required.  While this could be attributed to the user task being too simple, it leads P2's team to draw an alternate conclusion: \textit{``We stepped back and we thought well, these are non-expert users. Maybe we need to find ways to encourage them to interact with the data more. So yes, you could give them harder tasks, but what you could be doing is something more like surfacing bread crumbs to help somebody that's not an expert user start an investigation''}. Exposing unexpected usage patterns at the top level of the measurement hierarchy leads to additional insights about the application's overall design. At the more granular sub-component level, time spent and interaction measurements can facilitate the validation of hypotheses about how a system would be used, as illustrated by P2: \textit{``definitely it validated some assumptions we had about how the user would use the application in terms of where they spend their time, and we did see some iteration between the model explanation, the data transformation, data exploration components. That all lined up''}.  Likewise, the lower level measurements can expose unexpected patterns of use, or allow for key workflow components - those visited often, or used longer - to be readily identified.

\textbf{Effect of multi-faceted measurement}: Three organizers stated that their team appreciated the support for a multi-faceted measure because combining these different types of the signal allowed them to understand what sub-components work and what don't, and more importantly why. For instance, P4 mentioned they identified what sub-components really block the user flow by combining attitudinal and behavioral data: \textit{``We can see the impact of each component on the system...if I combine both of attitudinal and behavioral signal, we can identify which components are working well and which don't''}. She mentioned how her team's practice of checking total amount of time spent for every component scored low in attitudinal data helped them identify what needs to be done to improve their system: \textit{For this component [the component with a lower score on attitudinal data], the visualizations that we use to explain the model, so we are seeing that maybe our users just spend a lot of time there because they couldn't understand the visualization}. P10 pointed out a more specific insight that the multi-faceted measures lead to: \textit{``For model comparison, yeah. I mean, we look down here [behavioral 1, that shows each user's time spent in every sub-component] and we see only four people used it. We want to look at what people thought about it, and the attitudinal ratings on it are reasonably high. Only four people used it.''}. This allowed their team to conclude that they \textit{`` have to do something to make model comparison less hidden and more obvious to the users, to demonstrate that capability''}. The measure of use, paired with corresponding attitudinal data, can lead to deeper, more actionable insights.

In particular, our participants mentioned that the efficiency and effectiveness assessments performed at the component level allowed their teams to learn the specific parts of their systems that didn't work well, and those helped them to identify some break-through to be made. Some components presented very clear signal; \textit{``if you look at the attitudinal part where we evaluate the effectiveness and efficiency of the system...we can clearly see that model explanation technique we applied for classification has a large negative impact on our overall evaluation''}. P2 found even small rating differences between components lead to useful insights: \textit{``We saw that our regression analysis screen didn't do quite as well as our classification analysis screen. Although they're similar, it was enough that the differences in the functionality between the two produced a lower score. That was a little unexpected, but it was quickly identifiable from the results.''}. This shows that the approach used to measure and visualize the usability of the components can express larger trends, as well as smaller relative differences that can be just as important to an evaluation. 

It should be noted that the necessity of separate measures for perceived efficiency and perceived effectiveness is not clear given our observations. Questions were raised from Visus and TwoRavens as to the necessity of the split between efficiency and effectiveness in the attitudinal metrics. In almost all cases, the two are very highly correlated, pointing to the users not differentiating between the two concepts when responding to the survey.  P10 raises the issue: \textit{``That's not going to happen by chance, or there's no way that your system is equally effective and efficient. I think there has to be some re-framing of the questions. There has to be something different to separate the signal''}. However, Distil found the case where the effectiveness scores for model explanation components are observably lower than efficiency. P2 mentioned: \textit{``I think separating the visualization [into efficiency and effectiveness] is useful. It's more just making sure the user really things explicitly about the split between the two, would maybe get better results into that. But as I said, we still found signal in there that was interesting for us''} This disagreement between two parties raise the questions in the survey may need re-framing in order to make this measures make merit and worth pursuing.


\textbf{Effect of between-system comparisons}: One of the goals of our evaluation methodology is to provide first-class support for between-system comparison - both attitudinal and behavioral measures for each system are collected and made available in the analysis cards.  This comparison allows for the identification of the unique value a particular system delivers. P10 noted: textit{``Seeing how your system compares on certain dimensions to other systems helps you clarify how your system could be useful and is distinct from others.  Not everything is on a scale from better to worse, but rather, just seeing differences helps you think about what you might want to change or what you might want to keep the same.''}. This surfacing of distinctness is complimented by the ease in which patterns of similarity can be extracted across the evaluated systems. Given that the system components are all mapped to a defined hierarchy, participants were able to identify commonalities in component purposes and behaviors, and use that to understand parts of an AutoML workflow that are unfamiliar to test users vs. parts of the workflow that are poorly implemented or addressed by a given system. P2 states that \textit{``We could compare on some of the results that were a little surprising to us with the other systems and then see that the other systems had similar results here...so maybe that's just a result of the experiment population and not necessarily a problem with the application''}. One issue regarding between-system comparison was related to SUS score. SUS does provide a single metric to capture overall system usability, it seems that it is too coarse to derive any particular insight as we discussed in Section 2. The suggestion was made to possibly break it down into multiple scores, although those values would overlap with existing efficiency and effectiveness metrics, which would likely lead to confusion.

\subsection{Limitations, Discussion, and Future Work}
\label{sec:implications}

We discuss limitations of our methodology and directions we discovered for designing a better research methodology for evaluating systems for supporting exploratory model building process and visual analytics. 

One limitation that was consistently identified by participants was the absence of support for exploring results when they intend to ``fly through'' the data with and further inquiry. We characterize such directions of inquiry with the following three directions: \textit{Users} and \textit{Levels}.

Participants wished they can split end-users into subgroups depending on their behavioral patterns, and background knowledge about ML or visual analytics. The evaluation cards do include a user identifier as part of the effort visualizations (see Figure~\ref{fig:cards} (c)), but there is no way to see how the data for a user or group of users are distributed across other visualizations. P2 indicated that \textit{``at minimum, we would like to see some sort of a link between each of the individual users that we had and the usability metrics''}. Such a linking mechanism could also support the exploration of user types, information that is not available in the existing cards.  Three groups all indicated that some classification of users would help with interpreting results, as it would allow for users that are more experienced with machine learning or statistical analysis to be separated from those who are not.  P4 provides an example:\textit{``You can see users...that didn't spend a lot of time in model comparison and explanation. This could be because they don't understand machine learning in general.  We would like to be able to identify these users''}. 

Another aspect that participants mentioned they wished to have is the capability of exploring four categories of visualizations depending on the level and a specific component. To facilitate easy interpretation and reduce the learning curve, visualizations are currently presenting a static visualization of end-users' exploration behavior and attitudinal data. However, several participants indicated that Level 3 visualizations would be useful in more deeply understanding how users work with different sub-components while acknowledging that it would come at the cost of cross-system comparison.  P10 indicates that \textit{``you can gain some insight looking at L2, but you could also gain insight looking at L3. You can't compare across the systems looking at L3, but you could still look at it within your own system''}.  P2 expressed similar sentiments: \textit{``What would have been useful for us was to get logs for L3 so we could understand a little bit better what users were and were not doing with the application.  So less useful for between application, but more useful for us just trying to identify some usage patterns''}.

We envision that transforming the current evaluation categories into an interactive framework would require rigorous and in-depth design inquiry for uncovering system requirements. But we expect that such attempt may give us a useful lens that allows us to more deeply investigate a set of meaningful questions related to exploratory visual analytics, such as the relations between end-user background knowledge and their exploration patterns of a system, the relations between end-user system exploration patterns and their quality of outcomes, relations between the design of a particular component and the way people interact with the system, and more. In relation to the ``quality of outcome'', we did not define a ground-truth success metric in our tasks. The most common metric can be a model prediction accuracy, but we didn't use it because there can be numerous ways to determine the accuracy, and more importantly, we thought this metric is not a good proxy for measuring success in an exploratory task. However, we believe that establishing a more reliable proxy for measuring success in the exploratory task type would be an important future research agenda.


%% file: sections/06_Conclusion.tex
\section{Conclusions}

Our overarching goal was to better understand end-users' exploratory model building process through a novel evaluation methodology. Through the studies, we identified a set of insights for (1) developing AutoML systems that facilitate end-users' reasoning about their ML model's behavior and (2) performing more systematic evaluations which could lead to more specific and actionable insights than the current practice allows. As such, we hope our findings can motivate future research for defining a more systematic and rigorous approach for building and evaluating systems for supporting end-users' exploratory ML model building process.